\documentclass[nofootinbib,aps,pra,showpacs,superscriptaddress,preprint]{revtex4}
\usepackage{amsmath}
\usepackage{subfig}
\usepackage{graphicx}

\catcode`ð=\active
\defð{\u{g}}
\catcode`Ð=\active
\defÐ{\u{G}}
\catcode`Ý=\active
\defÝ{\. I}
\catcode`ö=\active
\defö{\"{o}}
\catcode`Ö=\active
\defÖ{\"O}
\catcode`ü=\active
\defü{\"{u}}
\catcode`Ü=\active
\defÜ{\"{U}}
\catcode`Þ=\active
\defÞ{\c{S}}
\catcode`þ=\active
\defþ{\c{s}}
\catcode`ý=\active
\defý{{\i}}
\catcode`ç=\active
\defç{\d{c}}
\catcode`Ç=\active
\defÇ{\d{C}}

\begin{document}

\title{Effective-Mass Dirac Equation for Woods-Saxon Potential: Scattering, Bound States and Resonances}
\author{\small Oktay Aydoðdu}
\email[E-mail: ]{oktaydogdu@gmail.com}\affiliation{Department of
Physics, Mersin University, 33343, Mersin,Turkey}
\author{\small Altuð Arda}
\email[E-mail: ]{arda@hacettepe.edu.tr}\affiliation{Department of
Physics Education, Hacettepe University, 06800, Ankara,Turkey}
\author{\small Ramazan Sever}
\email[E-mail: ]{sever@metu.edu.tr}\affiliation{Department of
Physics, Middle East Technical  University, 06531, Ankara,Turkey}

\begin{abstract}
Approximate scattering and bound state solutions of the
one-dimensional effective-mass Dirac equation with the Woods-Saxon
potential are obtained in terms of the hypergeometric-type
functions. Transmission and reflection coefficients are calculated
by using behavior of the wave functions at infinity. The same
analysis is done for the constant mass case. It is also pointed
out that our results are in agreement with those obtained in
literature. Meanwhile, an analytic expression is obtained for the
transmission resonance and observed that the expressions for bound
states and resonances are equal
for the energy values $E=\pm m$.\\
Keywords: Scattering, Bound State, Resonance, Dirac Equation,
Woods-Saxon Potential, Position-Dependent Mass
\end{abstract}
\pacs{03.65N, 03.65G, 03.65.Pm}

\maketitle

\newpage

\section{Introduction}

In order to get a complete information about quantum mechanical
systems, one should study bound and scattering states in the
presence of an external potential. Therefore, scattering problem
has become interesting topic in relativistic/non-relativistic
quantum mechanics. The problem has been studied within the
framework of group theoretical approach [1] and investigated for
the well-known potentials by applying different methods [2-18]. In
the case of non-relativistic scattering problem, it has been
showed that transmission and reflection coefficients take $1$ and
$0$, respectively, as external potential has well-behaved at
infinity for the zero energy limit [10-12]. However, reflection
coefficient goes to zero while transmission coefficient goes to
unity in the zero energy limit when external potential supports a
half-bound state. This situation was called as transmission
resonance by Bohm [13]. This phenomenon has been recently extended
to the Dirac particle in Refs. [3, 14, 15]. In the zero-momentum
limit, Dombey \textit{et. al} displayed that the bound-state
energy eigenvalue obtained for the Dirac particle in the presence
of the Woods-Saxon potential is related to the transmission
resonance appearing for a Dirac particle scattered by a potential
well. Recently, Villalba \textit{et. al} [6] have showed that
relation between the bound-state energy eigenvalues and
transmission resonances in view of the Woods-Saxon potential for
the Klein-Gordon particle are the same as obtained for the Dirac
particle [15].

On the other hand, solutions of the wave equations have, recently,
become interesting in the view of position-dependent mass (PDM)
formalism. Extensive applications of this formalism have been done
in different areas of physics such as condensed matter physics and
material science such as electronic properties of semiconductors
[19], quantum dots [20], quantum liquids \textit{etc.} [21-25]. In
recent years, the scattering problem has been extended to the case
where the mass depends on spatially coordinate [26-28]. Alhaidari
has, recently, investigated solution of the Dirac equation in view
of position-dependent mass for the Coulomb field [24]. In Ref.
[27], the authors have studied the relativistic scattering in the
Dirac equation by using the J-matrix method for the
position-dependent mass. Panella \textit{et. al} [28] have
obtained a new exact solution of the effective-mass Dirac equation
for the Woods-Saxon potential. The approximate solution of the
Dirac equation with PDM for the generalized Hulth\'{e}n potential
has been obtained by Peng \textit{et. al} [29]. Jia and co-workers
have extended $PT$-symmetric quantum mechanics for the Dirac
theory to the PDM formalism [30, 31].

In this work, we intend to solve the effective-mass Dirac equation
for the Woods-Saxon potential and investigate the scattering and
bound state solutions. We also study the transmission resonances
and give some results in the case of the low momentum limit.

The organization of this work is as follows. In Section II, we
give the one-dimensional Dirac equation for the case of
position-dependent mass and obtain a Schrödinger-like equation. In
Section III, we calculate the transmission and reflection
coefficients by analyzing the behavior of the wave functions at $x
\rightarrow \mp \infty$. To compare our results, we also calculate
the same coefficients for the case of constant mass and give our
results. In Section IV, we study the bound state problem for the
effective-mass Dirac equation and give the transmission resonance
and bound state equations for the low momentum limit. Conclusions
are given in Section V.

\section{Dirac Equation wýth Position Dependent Mass}
The relativistic free-particle Dirac equation ($\hbar=c=1$) is
written as [3, 15]
\begin{eqnarray}
\left[i\gamma^{\mu}\partial_{\mu}-m(x)\right]\psi(x)=0\,,
\end{eqnarray}
where we assume that the mass of the Dirac particle depends only
on one spatially coordinate $x$. Under the effect of an external
potential $V(x)$ and taking the gamma matrices $\gamma_{x}$ and
$\gamma_{0}$ as the Pauli matrices $i\sigma_{x}$ and $\sigma_{z}$,
respectively, the Dirac equation in one-dimension becomes
\begin{eqnarray}
\left\{\left(  \begin{array}{cc}
0 & 1 \\
1 & 0 \\
\end{array} \right)\frac{d}{dx}-[E-V(x)]\left(  \begin{array}{cc}
1 & 0 \\
0 & -1 \\
\end{array} \right)+m(x)\left(  \begin{array}{cc}
1 & 0 \\
0 & 1 \\
\end{array} \right)\right\}\left(  \begin{array}{c}
\varphi_{1}(x)\\
\varphi_{2}(x)\\
\end{array} \right)=0,
\end{eqnarray}
which gives the two following couple differential equations
\begin{eqnarray}
\frac{d\varphi_{1}(x)}{dx}&=&-\left[E-V(x)+m(x)\right]\varphi_{2}(x)\,,\\
\frac{d\varphi_{2}(x)}{dx}&=&\left[E-V(x)-m(x)\right]\varphi_{1}(x)\,.
\end{eqnarray}

The solutions can be more easily obtained by using a two-component
approach introduced by Flügge as [32]
\begin{eqnarray}
\phi(x)&=&\varphi_{1}(x)+i\varphi_{2}(x),\\
\chi(x)&=&\varphi_{1}(x)-i\varphi_{2}(x),
\end{eqnarray}
which leads
\begin{eqnarray}
\frac{d\phi(x)}{dx}&=&i[E-V(x)]\phi(x)-im(x)\chi(x),\\
\frac{d\chi(x)}{dx}&=&-i[E-V(x)]\chi(x)+im(x)\phi(x).
\end{eqnarray}

Eliminating $\chi(x)$ in Eq. (7) and inserting into Eq. (8) and
following the similar procedure for $\phi(x)$, we obtain two
uncoupled second-order differential equations for $\phi(x)$ and
$\chi(x)$, respectively
\begin{eqnarray}
\frac{d^2\phi(x)}{dx^2}&-&\frac{dm(x)/dx}{m(x)}\frac{d\phi(x)}{dx}+\bigg\{[E-V(x)]^2-m^2(x)+i\frac{dV(x)}{dx}
\nonumber\\&+&i[E-V(x)]\frac{dm(x)/dx}{m(x)}\bigg\}\phi(x)=0\,,
\end{eqnarray}
and
\begin{eqnarray}
\frac{d^2\chi(x)}{dx^2}&-&\frac{dm(x)/dx}{m(x)}\frac{d\chi(x)}{dx}+\bigg\{[E-V(x)]^2-m^2(x)-i\frac{dV(x)}{dx}
\nonumber\\&-&i[E-V(x)]\frac{dm(x)/dx}{m(x)}\bigg\}\chi(x)=0\,.
\end{eqnarray}
\section{Mass Function and Scattering State Solutions}
We assume that the mass of the Dirac particle depends on spatially
coordinate giving as
\begin{eqnarray}
m(x)=m_{0}+m_{1}f(x)\,,
\end{eqnarray}
where the function of $x$ as $f(x)=1/(1+e^{\alpha(|x|-L)})$. The
parameter $m_{0}$ will correspond to the rest mass of the particle
and $m_{1}$ is a real, positive, small parameter. The mass form
provides us to obtain the analytical results for the reflection
and transmission coefficients and also the bound state solutions
for the case of position-dependent mass and to analyze the results
for the case of constant mass. On the other hand, from Eq. (11),
it is easy to see that the ratio of the derivative of the mass to
the mass is proportional with the mass parameter $m_{1}$. So we
ignore the terms that contain the derivative of the mass in Eqs.
(9) and (10) for the case of $m_{1} \rightarrow 0$ [29]. Under
this assumption, Eqs. (9) and (10) become
\begin{equation}
\frac{d^2\phi(x)}{dx^2}+\bigg\{[E-V(x)]^2-m^2(x)+i\frac{dV(x)}{dx}\bigg\}\phi(x)=0\,,
\end{equation}
\begin{equation}
\frac{d^2\chi(x)}{dx^2}+\bigg\{[E-V(x)]^2-m^2(x)-i\frac{dV(x)}{dx}\bigg\}\chi(x)=0\,.
\end{equation}

We search the scattering states of the Dirac equation for the WS
potential barrier [3]
\begin{eqnarray}
V(x)=V_{0}f(x)\,,
\end{eqnarray}
where the function $f(x)$ is defined in Eq. (11) with $V_{0},
\alpha$ and $L$ are real parameters. This potential is one of the
most important potential models in quantum mechanics and has a
main role, as an internuclear potential, in the coupled-channels
calculations within the heavy-ion physics [33]. The nuclear
optical-model potential including the Woods-Saxon potential is
used to analyze the elastic scattering problem of nucleons and
heavy particles [34]. It is worth to say that we deal with a
potential form for $aL\gg1$. In this case, the potential form now
closely becomes a rectangular barrier.
\subsection{Solutions for $x<0$}
Using Eqs. (11) and (14) and defining a new variable
$y=(1+e^{-\alpha(x+L)})^{-1}$, Eq. (12) turns into
\begin{eqnarray}
y(1-y)\frac{d^2\phi_{L}(y)}{dy^2}+(1-2y)\frac{d\phi_{L}(y)}{dy}+\frac{1}{y(1-y)}\left\{C_{1}-C_{2}y+C_{3}y^2\right\}\phi_{L}(y)=0\,,
\end{eqnarray}
where
\begin{eqnarray}
C_{1}=\frac{E^2-m_{0}^2}{\alpha^2}\,\,\,;\,\,\,C_{2}=\frac{2V_{0}
E+2m_{0}m_{1}-iV_{0}\alpha
}{\alpha^2}\,\,\,;C_{3}=\frac{V_{0}^2-m_{1}^2-iV_{0}\alpha}{\alpha^2}\,.
\end{eqnarray}
In order to get a hypergeometric-type differential equation, we
offer a trial function $\phi_{L}(y)=y^{\mu}(1-y)^{\nu}f(y)$.
Substitution it into Eq. (15) leads to
\begin{eqnarray}
y(1-y)\frac{d^2f(y)}{dy^2}+[1+2\mu-2(\mu+\nu+1)y]\frac{df(y)}{dy}-(\mu+\nu+\sigma)(\mu+\nu-\sigma+1)f(y)=0\,,\nonumber\\
\end{eqnarray}
with
\begin{eqnarray}
\sigma=\frac{1}{2}+\sqrt{\left(\frac{iV_{0}}{\alpha}+\frac{1}{2}\right)^2+\frac{m_{1}^2}{\alpha^2}\,}\,\,;\nu=\frac{i}{\alpha}\sqrt{(E-V_{0})^2-(m_{0}+m_{1})^2\,}
\,\,;\mu=\frac{ik}{\alpha}\,;k=\sqrt{E^2-m_{0}^2\,}\,.\nonumber\\
\end{eqnarray}
Eq. (17) has a general solution [35]
\begin{eqnarray}
f(y)&=&L_{1}\,_2F_{1}(\mu+\nu+\sigma,\mu+\nu-\sigma+1,1+2\mu;y)
\nonumber\\&+&L_{2}y^{-2\mu}\,_2F_{1}(-\mu+\nu+\sigma,-\mu+\nu-\sigma+1,1-2\mu;y),
\end{eqnarray}
which gives
\begin{eqnarray}
\phi_{L}(y)&=&L_{1}y^{\mu}(1-y)^{\nu}\,_2F_{1}(a,b,c;y)
+L_{2}y^{-\mu}(1-y)^{\nu}\,_2F_{1}(a_{1},b_{1},c_{1};y),
\end{eqnarray}
where
\begin{eqnarray}
a&=&\mu+\nu+\sigma\,\,\,;\,\,\,\,\,\,\,\,\,\,\,\,\,\,\,a_{1}=-\mu+\nu+\sigma\,,\nonumber\\
b&=&\mu+\nu-\sigma+1\,\,\,;\,\,\,\,\,b_{1}=-\mu+\nu-\sigma+1\,,\nonumber\\
c&=&1+2\mu\,\,\,;\,\,\,\,\,\,\,\,\,\,\,\,\,\,\,\,\,\,\,\,\,\,\,\,c_{1}=1-2\mu\,.
\end{eqnarray}
\subsection{Solutions for $x>0$}
In this case, changing the variable $z=1/(1+e^{\alpha(x-L)})$
and inserting Eqs. (11) and (14) into Eq. (12), we get
\begin{eqnarray}
z(1-z)\frac{d^2\phi_{R}(z)}{dz^2}+(1-2z)\frac{d\phi_{R}(z)}{dz}+\frac{1}{z(1-z)}\left\{C_{4}-C_{5}z+C_{6}z^2\right\}\phi_{R}(z)=0\,,
\end{eqnarray}
with
\begin{eqnarray}
C_{4}=\frac{E^2-m_{0}^2}{\alpha^2}\,\,\,;\,\,\,C_{5}=\frac{2
E+2m_{0}m_{1}+iV_{0}\alpha
}{\alpha^2}\,\,\,;C_{6}=\frac{V_{0}^2-m_{1}^2+iV_{0}\alpha}{\alpha^2}\,.
\end{eqnarray}
Defining a wave function of the form
$\phi_{R}(z)=z^{\tau}(1-z)^{\gamma}h(z)$ in Eq. (22) gives a
hypergeometric-type equation [35]
\begin{eqnarray}
z(1-z)\frac{d^2h(z)}{dz^2}+[1+2\tau-2(\tau+\gamma+1)z]\frac{dh(z)}{dz}-(\tau+\gamma+\delta)(\tau+\gamma-\delta+1)h(z)=0\,,\nonumber\\
\end{eqnarray}
where we have obtained as $\tau=\mu$\,,$\gamma=\nu$ and
$\delta=\frac{1}{2}+\sqrt{\left(\frac{iV_{0}}{\alpha}-\frac{1}{2}\right)^2+\frac{m_{1}^2}{\alpha^2}\,}$\,.
Eq. (24) has a solution in terms of hypergeometric functions [35]
\begin{eqnarray}
h(z)&=&R_{1}\,_2F_{1}(\mu+\nu+\delta,\mu+\nu-\delta+1,1+2\mu;z)
\nonumber\\&+&R_{2}z^{-2\mu}\,_2F_{1}(-\mu+\nu+\delta,-\mu+\nu-\delta+1,1-2\mu;z),
\end{eqnarray}
which gives the following complete solution for $x>0$
\begin{eqnarray}
\phi_{R}(z)&=&R_{1}z^{\mu}(1-z)^{\nu}\,_2F_{1}(a_{3},b_{3},c_{3};z)
+R_{2}z^{-\mu}(1-z)^{\nu}\,_2F_{1}(a_{2},b_{2},c_{2};z),
\end{eqnarray}
where
\begin{eqnarray}
a_{2}&=&-\mu+\nu+\delta\,\,\,;\,\,\,\,\,\,\,\,\,\,\,\,\,\,\,a_{3}=\mu+\nu+\delta\,,\nonumber\\
b_{2}&=&-\mu+\nu-\delta+1\,\,\,;\,\,\,\,\,b_{3}=\mu+\nu-\delta+1\,,\nonumber\\
c_{2}&=&1-2\mu\,\,\,;\,\,\,\,\,\,\,\,\,\,\,\,\,\,\,\,\,\,\,\,\,\,\,\,\,\,\,\,c_{3}=1+2\mu\,.
\end{eqnarray}
If $x \rightarrow \infty$, $z \rightarrow 0$, then $(1-z)^{\nu} \rightarrow 1$ and
$z^{\mu} \rightarrow e^{-\alpha\mu(x-L)}$. Thus, we obtain the following right solution in this limit
\begin{equation}
\phi_{R}(x)\approx R_{1}e^{-ik(x-L)}+R_{2}e^{ik(x-L)},
\end{equation}
where we have used the following property of the hypergeometric functions:
$\,_2F_{1}(\xi_{1},\xi_{2},\xi_{3};t)\xrightarrow[t \to 0]{}1$\,.
In order to get a plane wave coming from the left to the right, we set $R_1=0$. Consequently, the right solution becomes
\begin{equation}
\phi_{R}(z)=R_{2}z^{-\mu}(1-z)^{\nu}\,_2F_{1}(a_{2},b_{2},c_{2};z).
\end{equation}

\subsection{Reflection and Transmission Coefficients}
Let us now study the behavior of the wave functions $\phi_{L}(y)$
and $\phi_{R}(z)$ at infinity to obtain the reflection and
transmission coefficients. In the limit $x \rightarrow -\infty$,
$y \rightarrow 0$, $(1-y)^{\nu} \rightarrow 1$ and $y^{\mu}
\rightarrow e^{\alpha\mu(x+L)}$ and in the limit $x \rightarrow
\infty$, $z \rightarrow 0$, $(1-z)^{\nu} \rightarrow 1$ and
$z^{\mu} \rightarrow e^{-\alpha\mu(x-L)}$ as well (with
$\,_2F_{1}(\xi_{1},\xi_{2},\xi_{3};t)\xrightarrow[t \to 0]{}1$\,),
we have the wave functions, respectively,
\begin{eqnarray}
&&\phi_{L}(x)\xrightarrow[x \,\to\,
-\infty]{}L_{1}e^{ik(x+L)}+L_{2}e^{-ik(x+L)}\,,\\
&&\phi_{R}(x)\xrightarrow[x \,\to\, \infty]{}R_{2}e^{ik(x-L)}\,.
\end{eqnarray}

In order to get the electrical current density for the
one-dimensional Dirac equation defined by
\begin{eqnarray}
j=\frac{1}{2}\big[\left|\phi(x)\right|^2-\left|\chi(x)\right|^2\big],
\end{eqnarray}
we need to insert Eqs. (30) and (31) into Eq. (7) which gives
$\chi_{L}(x)$ and $\chi_{R}(x)$, respectively,
\begin{eqnarray}
\chi_{L}(x)&=&\frac{1}{m(x)}\left[(E-k)L_{1}e^{ik(x+L)}+(E+k)L_{2}e^{-ik(x+L)}\right],\\
\chi_{R}(x)&=&\left(\frac{E-k}{m(x)}\right)R_{2}e^{ik(x+L)}.
\end{eqnarray}
The current in Eq. (32) can be written as $j_{L}=j_{in}-j_{refl}$
in the limit $x \rightarrow -\infty$ where $j_{in}$ is the
incident and $j_{refl}$ is the reflected current. Similarly as $x
\rightarrow \infty$ the current is $j_{R}=j_{trans}$ where
$j_{trans}$ is the transmitted current. Inserting Eqs. (30), (31),
(33) and (34) into Eq. (32), we find the reflection and
transmission coefficients, respectively, as
\begin{eqnarray}
R&=&\frac{\left(E+k\right)}{\left(E-k\right)}\frac{|L_{2}|^2}{|L_{1}|^2}\,,\\
T&=&\frac{|R_{2}|^2}{|L_{1}|^2}\,.
\end{eqnarray}
We can find more explicit expressions for the above coefficients
by using the continuity conditio of the wave function at $x=0$.
For the limit $x \rightarrow 0$, we have $y \rightarrow 1$ and
$\,_2F_{1}(\xi_{1},\xi_{2},\xi_{3};t)\xrightarrow[t \to 0]{}1$,
the wave function $\phi_{L}(x)$ ($aL\gg0$)
\begin{eqnarray}
\phi_{L}(x)\xrightarrow[x \to 0]{}\left[L_{1}S_{1}e^{-\alpha\nu
L}+L_{2}S_{2}e^{-\alpha\nu L}\right]e^{-\alpha\nu
x}+\left[L_{1}S_{3}e^{\alpha\nu L}+L_{2}S_{4}e^{\alpha\nu
L}\right]e^{\alpha\nu x}\,,
\end{eqnarray}
where we have used the following identity of the hypergeometric
functions [35]
\begin{eqnarray}
_2F_{1}(\xi_{1},\xi_{2},\xi_{3};t)&=&\frac{\Gamma(\xi_{3})\Gamma(\xi_{3}-\xi_{2}-\xi_{1})}
{\Gamma(\xi_{3}-\xi_{1})\Gamma(\xi_{3}-\xi_{2})}\,_2F_{1}(\xi_{1},\xi_{2},\xi_{1}+\xi_{2}-\xi_{3};1-t)\nonumber\\
&+&(1-t)^{\xi_{3}-\xi_{2}-\xi_{1}}\frac{\Gamma(\xi_{3})\Gamma(\xi_{1}+\xi_{2}-\xi_{3})}
{\Gamma(\xi_{1})\Gamma(\xi_{2})}\,_2F_{1}(\xi_{3}-\xi_{1},\xi_{3}-\xi_{2},\xi_{3}-\xi_{2}-\xi_{1}+1;1-t)\,.\nonumber\\
\end{eqnarray}
The abbreviations in Eq. (37) are
\begin{eqnarray}
S_{1}&=&\frac{\Gamma(c)\Gamma(c-a-b)}{\Gamma(c-a)\Gamma(c-b)}\,\,\,;\,\,\,\,
S_{2}=\frac{\Gamma(c_{1})\Gamma(c_{1}-a_{1}-b_{1})}{\Gamma(c_{1}-a_{1})\Gamma(c_{1}-b_{1})}\,,\nonumber\\
S_{3}&=&\frac{\Gamma(c)\Gamma(a+b-c)}{\Gamma(a)\Gamma(b)}\,\,\,;\,\,\,\,
S_{4}=\frac{\Gamma(c_{1})\Gamma(a_{1}+b_{1}-c_{1})}{\Gamma(a_{1})\Gamma(b_{1})}\,.
\end{eqnarray}
We have $z \rightarrow 1$ for the same limit $x \rightarrow 0$, so
we write $\phi_{R}(x)$ as
\begin{eqnarray}
\phi_{R}(x)\xrightarrow[x \to 0]{}
R_{2}\left[S_{5}e^{\alpha\nu(x-L)}+S_{6}e^{-\alpha\nu(x-
L)}\right]\,,
\end{eqnarray}
where
\begin{eqnarray}
S_{5}=\frac{\Gamma(c_{2})\Gamma(c_{2}-a_{2}-b_{2})}{\Gamma(c_{2}-a_{2})\Gamma(c_{2}-b_{2})}\,\,\,;\,\,\,\,
S_{6}=\frac{\Gamma(c_{2})\Gamma(a_{2}+b_{2}-c_{2})}{\Gamma(a_{2})\Gamma(b_{2})}\,.
\end{eqnarray}
Finally, from matching the wave functions in Eqs. (37) and (40),
we obtain
\begin{eqnarray}
\frac{L_{2}}{L_{1}}&=&\frac{e^{-4\alpha\nu
L}S_{1}S_{5}-S_{3}S_{6}}{S_{4}S_{6}-e^{-4\alpha\nu
L}S_{2}S_{5}}\,,
\end{eqnarray}
and
\begin{eqnarray}
\frac{R_{2}}{L_{1}}&=&\frac{e^{-2\alpha\nu
L}[S_{1}S_{4}-S_{3}S_{2}]}{S_{4}S_{6}-e^{-4\alpha\nu
L}S_{2}S_{5}}\,.
\end{eqnarray}

In Fig. 1, it is seen that the transmission and reflection
coefficients oscillate between the values zero and one and satisfy
the condition $R+T=1$ for the case of constant and PDM as well.
The oscillations appear in the range of $m_0<E<2m_0$ while $T=0\,
(R=1)$ as $E<m_0$ and $T=1\,(R=0)$ as $E>2m_0$. The effect of the
PDM is just to shift the picks to the left. We present the
variation of the transmission coefficient with respect to the
parameter $V_{0}$ in the case of PDM in Fig. 2. We also plot the
same variation for the constant mass. The transmission coefficient
goes to zero with increasingly high of potential barrier and
exactly zero for the values in the range $0.4<V_{0}<1.2$. The
coefficient $T$ starts to oscillate and does not take zero-value
for $V_{0}>1.2$ where the upper value of the oscillation for $T$
is one. Fig. 2 shows also the effect of PDM on the dependence of
$T$ on potential parameter $V_{0}$ and that this effect is very
weak. Fig. 3 shows the dependence of the transmission coefficient
on the potential parameters $\alpha$ (left plot) and $L$ (right
plot) in the case of PDM in view of the varying particle energy.
In both of the plots, $T$ oscillates between the values zero and
one as in the case of constant mass. It is seen that the frequency
of the oscillations increases while the parameter $L$ increases.
In Fig. 4, we show the effects of the potential parameters
$\alpha$ (left plot) and $L$ (right plot) on the variation of the
transmission coefficient with varying potential parameter $V_{0}$.
In these figures, coefficient of transmission is exactly zero
within the range of $m_0<V_{0}<3m_0$ and continue to oscillate out
of this range.

One interesting point is the so-called "transmission resonances"
appearing especially in relativistic domain [3, 6, 14, 15]. Within our
present formalism, considering Eq. (42), the transmission resonances ($R=0$, $T=1$) occur when
\begin{eqnarray}
e^{-4\alpha\nu L}S_1S_5-S_3S_6=0\,.
\end{eqnarray}

From Figs. 1 and 2, we see that the Dirac particle has
transmission resonances in the case of PDM like the case of
constant mass. In both of figures, we observe that the effect of
mass depending on coordinate is to shift the picks to the left.
Fig. 3 shows that, as in the case of constant mass, transmission
resonances appear. From the left plot of Fig. 3, one can observe
that the width of the resonance peaks decreases and the number of
the transmission resonances remains the same while the parameter
$\alpha$ decreases. In addition, it is seen from the left plot of
the Fig. 3 that the first resonance peak appears at smaller values
of the Dirac particle's energy in the presence of the small value
of the parameter $\alpha$. From the right panel of the Fig. 3, one
observes that number of the resonance peaks decreases while the
width of the resonance peaks increases with decreasing the
parameter $L$. From Fig. 4, we see that transmission resonances
could be observed with varying $V_{0}$ in the case of PDM and
width of the resonance peaks decreases while the parameter
$\alpha$ becomes smaller but it increases as the parameter $L$
decreases.
\section{Bound State Solutions}
We tend to find the bound states for the Woods-Saxon potential
well which means $V_{0} \rightarrow -V_{0}$ in Eq. (14).
\subsection{Solutions for $x<0$}
In order to get a complete solution for this region, we use a new
variable $y=[1+e^{-\alpha(x+L)}]^{-1}$ in Eq. (12) and take into
account $V_{0} \rightarrow -V_{0}$, we have
\begin{eqnarray}
y(1-y)\frac{d^2\phi_{L}(y)}{dy^2}+(1-2y)\frac{d\phi_{L}(y)}{dy}+\frac{1}{y(1-y)}\left\{C'_{1}-C'_{2}y+C'_{3}y^2\right\}\phi_{L}(y)=0\,,
\end{eqnarray}
where
\begin{eqnarray}
\underbrace{C'_{1}=\frac{E^2-m_{0}^2}{\alpha^2}}_{C'_{1}=C_{1}}\,\,\,;\,\,\,\underbrace{C'_{2}=\frac{-2V_{0}
E+2m_{0}m_{1}+iV_{0}\alpha }{\alpha^2}}_{C'_{2}\xrightarrow[V_{0}
\to -V_{0}]{} C_{2}
}\,\,\,;\underbrace{C'_{3}=\frac{V_{0}^2-m_{1}^2+iV_{0}\alpha}{\alpha^2}}_{C'_{3}\xrightarrow[V_{0}
\to -V_{0}]{} C_{3}}\,.
\end{eqnarray}
Taking a trial wave function as
$\phi_{L}(y)=y^{\mu'}(1-y)^{\nu'}g(y)$ and inserting it into Eq.
(45) we obtain
\begin{eqnarray}
y(1-y)\frac{d^2g(y)}{dy^2}+[1+2\mu'-2(\mu'+\nu'+1)y]\frac{dg(y)}{dy}-(\mu'+\nu'+\sigma')(\mu'+\nu'-\sigma'+1)g(y)=0\,,\nonumber\\
\end{eqnarray}
with
\begin{eqnarray}
\sigma'=\frac{1}{2}+\sqrt{\left(\frac{-iV_{0}}{\alpha}+\frac{1}{2}\right)^2+\frac{m_{1}^2}{\alpha^2}\,}\,\,;\nu'=\frac{i}{\alpha}\sqrt{(E+V_{0})^2-(m_{0}+m_{1})^2\,}
\,\,;\mu'=-\frac{1}{\alpha}\,\sqrt{m_{0}^2-E^2\,}\,.\nonumber\\
\end{eqnarray}
The solution of Eq. (47) is written in terms of the hypergeometric
type functions [35]
\begin{eqnarray}
\phi_{L}(y)&=&L_{3}y^{\mu'}(1-y)^{\nu'}\,_2F_{1}(a',b',c';y)
+L_{4}y^{-\mu'}(1-y)^{\nu'}\,_2F_{1}(a'_{1},b'_{1},c'_{1};y)
\end{eqnarray}
where
\begin{eqnarray}
a'&=&\mu'+\nu'+\sigma'\,\,\,;\,\,\,\,\,\,\,\,\,\,\,\,\,\,\,a'_{1}=-\mu'+\nu'+\sigma'\,,\nonumber\\
b'&=&\mu'+\nu'-\sigma'+1\,\,\,;\,\,\,\,\,b'_{1}=-\mu'+\nu'-\sigma'+1\,,\nonumber\\
c'&=&1+2\mu'\,\,\,;\,\,\,\,\,\,\,\,\,\,\,\,\,\,\,\,\,\,\,\,\,\,\,\,\,\,c'_{1}=1-2\mu'\,.
\end{eqnarray}
\subsection{Solutions for $x>0$}
Inserting the potential function
\begin{eqnarray}
V(x)=-\frac{V_{0}}{1+e^{\alpha(x-L)}}\,,
\end{eqnarray}
into Eq. (12) and using the variable $z=1/[1+e^{\alpha(x-L)}]$, we
get
\begin{eqnarray}
z(1-z)\frac{d^2\phi_{R}(z)}{dz^2}+(1-2z)\frac{d\phi_{R}(z)}{dz}+\frac{1}{z(1-z)}\left\{C'_{4}-C'_{5}z+C'_{6}z^2\right\}\phi_{R}(z)=0\,,
\end{eqnarray}
with
\begin{eqnarray}
\underbrace{C'_{4}=\frac{E^2-m_{0}^2}{\alpha^2}}_{C'_{4}=C_{4}}\,\,\,;\,\,\,\underbrace{C'_{5}=\frac{-2V_{0}
E+2m_{0}m_{1}-iV_{0}\alpha }{\alpha^2}}_{C'_{5}\xrightarrow[V_{0}
\to -V_{0}]{} C_{5}
}\,\,\,;\underbrace{C_{6}=\frac{V_{0}^2-m_{1}^2-iV_{0}\alpha}{\alpha^2}}_{C'_{6}\xrightarrow[V_{0}
\to -V_{0}]{} C_{6}}\,.
\end{eqnarray}
Taking a wave function of the form
$\phi_{R}(z)=z^{\mu'}(1-z)^{\nu'}w(z)$ in Eq. (52) gives a
hypergeometric-type equation [35]
\begin{eqnarray}
z(1-z)\frac{d^2w(z)}{dz^2}+[1+2\mu'-2(\mu'+\nu'+1)z]\frac{dw(z)}{dz}-(\mu'+\nu'+\delta')(\mu'+\nu'-\delta'+1)w(z)=0\,,\nonumber\\
\end{eqnarray}
where
$\delta'=\frac{1}{2}+\sqrt{\left(\frac{-iV_{0}}{\alpha}-\frac{1}{2}\right)^2+\frac{m_{1}^2}{\alpha^2}\,}$\,.
The general solution of Eq. (54) is written in terms of
hypergeometric functions as follow
\begin{eqnarray}
w(z)&=&R_{3}\,_2F_{1}(\mu'+\nu'+\delta',\mu'+\nu'-\delta'+1,1+2\mu';z)
\nonumber\\&+&R_{4}z^{-2\mu'}\,_2F_{1}(-\mu'+\nu'+\delta',-\mu'+\nu'-\delta'+1,1-2\mu';z),
\end{eqnarray}
and the whole solution for $x>0$ is
\begin{eqnarray}
\phi_{R}(z)&=&R_{3}z^{\mu'}(1-z)^{\nu'}\,_2F_{1}(a'_{3},b'_{3},c'_{3};z)
+R_{4}z^{-\mu'}(1-z)^{\nu'}\,_2F_{1}(a'_{2},b'_{2},c'_{2};z)
\end{eqnarray}
where
\begin{eqnarray}
a'_{2}&=&-\mu'+\nu'+\delta'\,\,\,;\,\,\,\,\,\,\,\,\,\,\,\,\,\,\,a'_{3}=\mu'+\nu'+\delta'\,,\nonumber\\
b'_{2}&=&-\mu'+\nu'-\delta'+1\,\,\,;\,\,\,\,\,b'_{3}=\mu'+\nu'-\delta'+1\,,\nonumber\\
c'_{2}&=&1-2\mu'\,\,\,;\,\,\,\,\,\,\,\,\,\,\,\,\,\,\,\,\,\,\,\,\,\,\,\,\,\,\,\,\,\,c'_{3}=1+2\mu'\,.
\end{eqnarray}
Let us now extract the solutions given in Eqs. (49) and (56) in
the limit $x \rightarrow \pm\infty$ to obtain the bound state wave
function. Because of $y \rightarrow 0$ and $z \rightarrow 0$ as
well for the limit $x \rightarrow \mp\infty$, we write $\phi_{L}$
in Eq. (49) and $\phi_{R}$ in Eq. (56), respectively
\begin{eqnarray}
\phi_{L}(y)\xrightarrow[x \to -\infty]{}
L_{3}y^{\mu'}(1-y)^{\nu'}\,_2F_{1}(a',b',c';y)\,,\\
\phi_{R}(z)\xrightarrow[x \to \infty]{}
R_{3}z^{\mu'}(1-z)^{\nu'}\,_2F_{1}(a'_{3},b'_{3},c'_{3};z)\,
\end{eqnarray}
where we set $L_{4}=R_{4}=0$ for obtaining the bound state
eigenfunctions. In order to study the behavior of the solution at
$x=0$, we need the property of the hypergeometric functions given
in Eq. (38). Recalling that for $x \rightarrow 0$, $y$, $z
\rightarrow 1$ and $(1-y)^{\nu'}\approx e^{-\alpha\nu'(x+L)}$
while $(1-z)^{\nu'}\approx e^{\alpha\nu'(x-L)}$ and using Eq.
(38), we write the wave functions
\begin{eqnarray}
\phi_{L}(x)\xrightarrow[x \to 0]{}
L_{3}\left[S'_{1}e^{-\alpha\nu'(x+L)}+S'_{3}e^{\alpha\nu'(x+L)}\right]\,,\\
\phi_{R}(x)\xrightarrow[x \to 0]{}
R_{3}\left[S'_{4}e^{\alpha\nu'(x-L)}+S'_{2}e^{-\alpha\nu'(x-L)}\right]\,
\end{eqnarray}
where
\begin{eqnarray}
S'_{1}&=&\frac{\Gamma(c')\Gamma(c'-a'-b')}{\Gamma(c'-a')\Gamma(c'-b')};S'_{2}=\frac{\Gamma(c'_3)\Gamma(a'_3+b'_3-c'_3)}{\Gamma(a'_3)\Gamma(b'_3)}\,,\nonumber\\
S'_{3}&=&\frac{\Gamma(c')\Gamma(a'+b'-c')}{\Gamma(a')\Gamma(b')};S'_{4}=\frac{\Gamma(c'_3)\Gamma(c'_3-a'_3-b'_3)}{\Gamma(c'_3-a'_3)\Gamma(c'_3-b'_3)}\,.
\end{eqnarray}

Matching the functions given in Eqs. (60) and (61) at $x=0$
requiring the continuity of the wave function, comparing the
coefficients of $e^{\pm\alpha\nu'x}$ and setting the coefficients
determinant to zero, we obtain the following eigenvalue condition
for the Woods-Saxon potential well
\begin{eqnarray}
f[\alpha,V_{0},m_0,m_1,E]=S'_{2}S'_{3}-S'_{1}S'_{4}e^{-4\alpha\nu'L}=0\,.
\end{eqnarray}
The above expression can be solved numerically and the energy
eigenvalues $E$ could be obtained by setting
Re$[f[\alpha,V_{0},m_0,m_1,E]]=0$ and also
Im$[f[\alpha,V_{0},m_0,m_1,E]]=0$ since
$f[\alpha,V_{0},m_0,m_1,E]=0$ is complex. We search the numerical
energy values for the interval $m-V_{0}\leq E \leq m$, since we
are interested in bound states. Figs. 5 and 6 show the real energy
eigenvalues for the PDM and constant mass cases, respectively. The
eigenvalues which are shown with arrows are the points on the
$E$-axis where the Re$[f[\alpha,V_{0},m_0,m_1,E]]$-curve (solid
line) and Im$[f[\alpha,V_{0},m_0,m_1,E]]$-curve (dotted line)
cross. From the Figs. 5 and 6, one can see that the number of
bound states in the case of PDM increases relative to the constant
mass case.

Finally, let us study our results in the case of low momentum
limit which has become an attractive topic especially in the
relativistic domain [3, 6, 14, 15]. In this limit, the Dirac
equation has two distinct states: One has the energy $E=m$
corresponding to particle state and the other one has the energy
$E=-m$ corresponding to anti-particle state where $m$ is the
particle mass [3].

Firstly, we investigate the resonance equation for the values of
$E=\pm m$. For this case, Eq. (44) becomes
\begin{eqnarray}
e^{-4\alpha\nu
L}\frac{\Gamma(c)\Gamma(c-a-b)}{\Gamma(c-a)\Gamma(c-b)}\frac{\Gamma(c_2)\Gamma(c_2-a_2-b_2)}{\Gamma(c_2-a_2)\Gamma(c_2-b_2)}=
\frac{\Gamma(c)\Gamma(a+b-c)}{\Gamma(a)\Gamma(b)}\frac{\Gamma(c_2)\Gamma(a_2+b_2-c_2)}{\Gamma(a_2)\Gamma(b_2)}\,,
\end{eqnarray}
where the arguments are obtained from Eqs. (21) and (27) as
\begin{eqnarray}
a&=&\nu(E \rightarrow \pm m)+\sigma\,\,\,;\,\,\,\,\,\,\,\,\,\,\,\,\,\,a_{2}=\nu(E \rightarrow \pm m)+\delta\,,\nonumber\\
b&=&\nu(E \rightarrow \pm m)-\sigma+1\,\,\,;\,\,\,\,\,b_{2}=\nu(E \rightarrow \pm m)-\delta+1\,,\nonumber\\
c&=&1\,\,\,;\,\,\,\,\,\,\,\,\,\,\,\,\,\,\,\,\,\,\,\,\,\,\,\,\,\,\,\,\,\,\,\,\,\,\,\,\,\,\,\,\,\,\,\,\,\,\,\,\,\,\,\,\,\,\,c_{2}=1\,.
\end{eqnarray}
Secondly, we write the bound state equation given in Eq. (63) for
$E=\pm m$ giving
\begin{eqnarray}
e^{-4\alpha\nu'
L}\frac{\Gamma(c')\Gamma(c'-a'-b')}{\Gamma(c'-a')\Gamma(c'-b')}\frac{\Gamma(c'_3)\Gamma(c'_3-a'_3-b'_3)}{\Gamma(c'_3-a'_3)\Gamma(c'_3-b'_3)}=
\frac{\Gamma(c')\Gamma(a'+b'-c')}{\Gamma(a')\Gamma(b')}\frac{\Gamma(c'_3)\Gamma(a'_3+b'_3-c'_3)}{\Gamma(a'_3)\Gamma(b'_3)}\,,\nonumber\\
\end{eqnarray}
where the arguments could be given from Eqs. (50) and (57) as
\begin{eqnarray}
a'&=&\nu'(E \rightarrow \pm m)+\sigma'\,\,\,;\,\,\,\,\,\,\,\,\,\,\,\,\,\,\,a'_{3}=\nu'(E \rightarrow \pm m)+\delta'\,,\nonumber\\
b'&=&\nu'(E \rightarrow \pm m)-\sigma'+1\,\,\,;\,\,\,\,\,b'_{3}=\nu'(E \rightarrow \pm m)-\delta'+1\,,\nonumber\\
c'&=&1\,\,\,;\,\,\,\,\,\,\,\,\,\,\,\,\,\,\,\,\,\,\,\,\,\,\,\,\,\,\,\,\,\,\,\,\,\,\,\,\,\,\,\,\,\,\,\,\,\,\,\,\,\,\,\,\,\,\,\,\,\,c'_{3}=1\,.
\end{eqnarray}
Because of the substituting $V_{0} \rightarrow -V_{0}$, we see
that $\nu'=\nu, \sigma'=\sigma$ and $\delta'=\delta$, so
\begin{eqnarray}
a'&=&a\,\,\,;\,\,\,\,\,a'_{3}=a_2\,,\nonumber\\
b'&=&b\,\,\,;\,\,\,\,\,b'_{3}=b_2\,,\nonumber\\
c'&=&c\,\,\,;\,\,\,\,\,c'_{3}=c_2\,.
\end{eqnarray}
which means that the resonance and bound state equations are equal
for the low momentum limit in both of the constant and
position-dependent mass cases. This result supports the one in
Ref. [3] which declares that the conditions for the tunnelling
without reflection trough a potential barrier $V(x)$ of a Dirac
particle with small momentum (resonances) and supporting of the
potential well $-V(x)$ a bound state energy $E=\pm m$ called
supercriticality are the same. It should be noted that the
transmission resonance of particles when they scatter off
potential barriers is equivalent to the one of anti-particles when
scattering off potential wells [6, 14, 15] and tunnelling of a
Dirac particle trough a potential barrier is strongly related to
the Klein paradox [14, 15].
\section{Conclusions}
We have approximately solved the scattering and bound state
problems in one-dimensional effective-mass Dirac equation for the
Woods-Saxon potential and studied the problem by using an
approximation in the mass distribution given as $m_{1} \rightarrow
0$. Using this approximation, we have found reflection and
transmission coefficients by analyzing the behavior of the
functions $\phi_{L}(x)$ ($x<0$) and $\phi_{R}(x)$ ($x>0$) at
$x\rightarrow \mp\infty$ within the framework of
position-dependent mass formalism. It has been observed that the
coefficients oscillate for the range where $E\gg m_{0}$ and $V_{0}
\gg m_{0}$ for both cases in which $T$ and $R$ change with energy
and with potential parameter, respectively. The unitarity
condition has also been checked numerically in the case of
position-dependent mass. We have pointed out that the results for
the case of constant mass are similar with the ones obtained in
the literature [3]. Meanwhile, the energy eigenvalue equation has
been found by using the wave function obtained by imposing the
boundary condition of a bound state. We have also studied the
effect of the mass varying with coordinate on transmission
resonances and on the results which are obtained for the low
momentum limit ($E=\pm m$). The relation between the transmission
resonance and bound-state energy eigenvalues has been presented in
the presence of the position dependent mass case.

\section{Acknowledgments}
This research was partially supported by the Scientific and
Technical Research Council of Turkey.

\newpage

\begin{figure}
\centering
\includegraphics[height=2in, width=4in, angle=0]{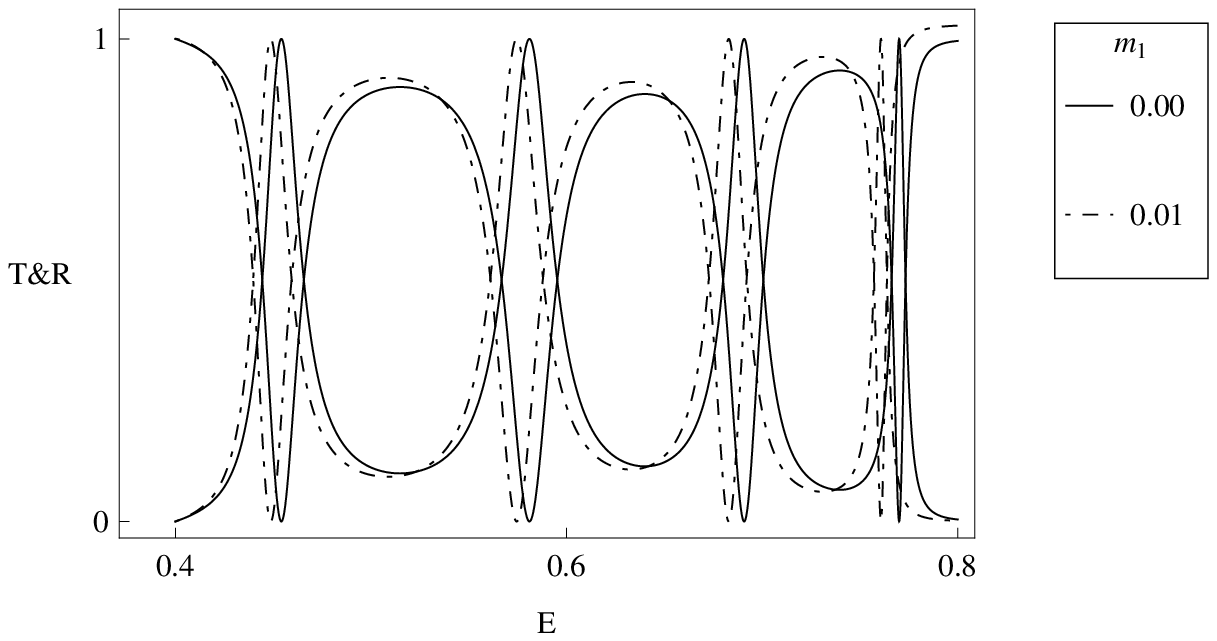}
\caption{transmission and reflection coefficients versus energy
with $m_{0}=0.4, L=10, \alpha=5, V_{0}=1.2$.}
\end{figure}

\begin{figure}
\centering
\includegraphics[height=2in, width=4in, angle=0]{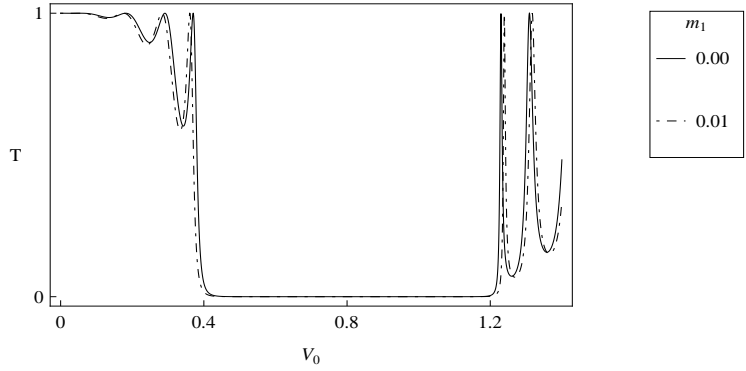}
\caption{coefficient of transmission versus potential parameter
$V_{0}$ with $m_{0}=0.4, L=10, \alpha=5, E=0.8$.}
\end{figure}

\begin{figure}
\centering \subfloat{
\includegraphics[width=2.8in]{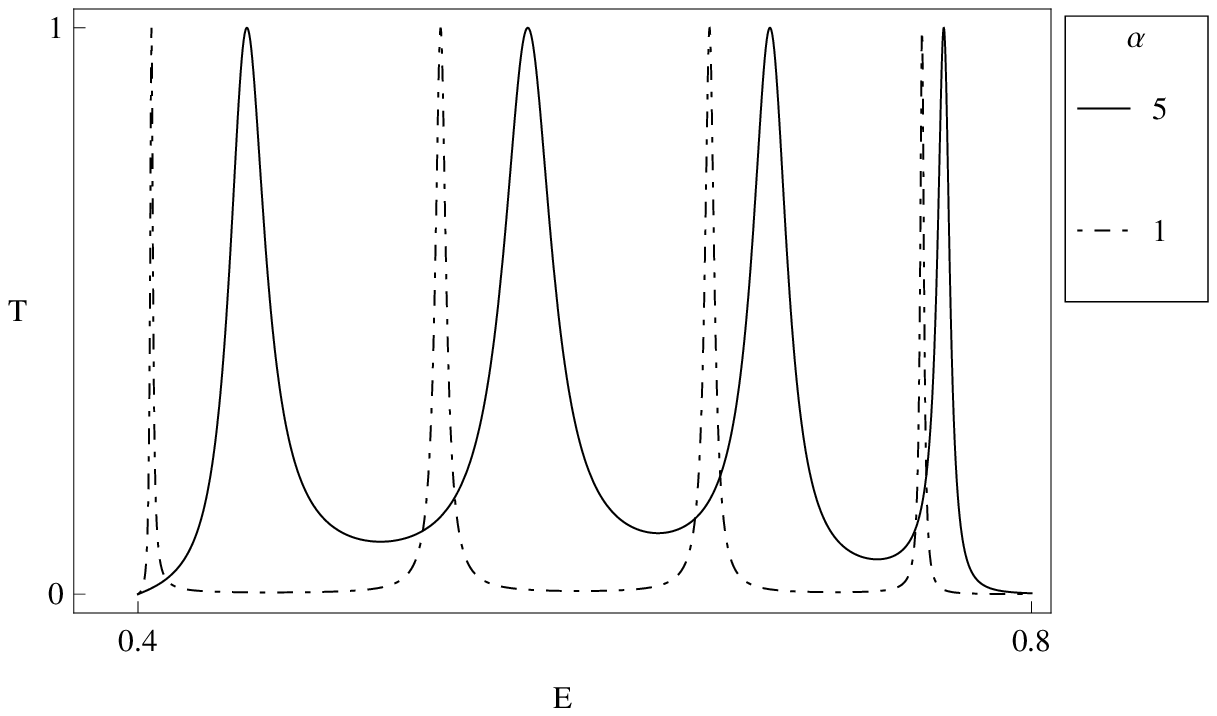}}
\hspace{0.1\linewidth} \subfloat{
\includegraphics[width=2.8in]{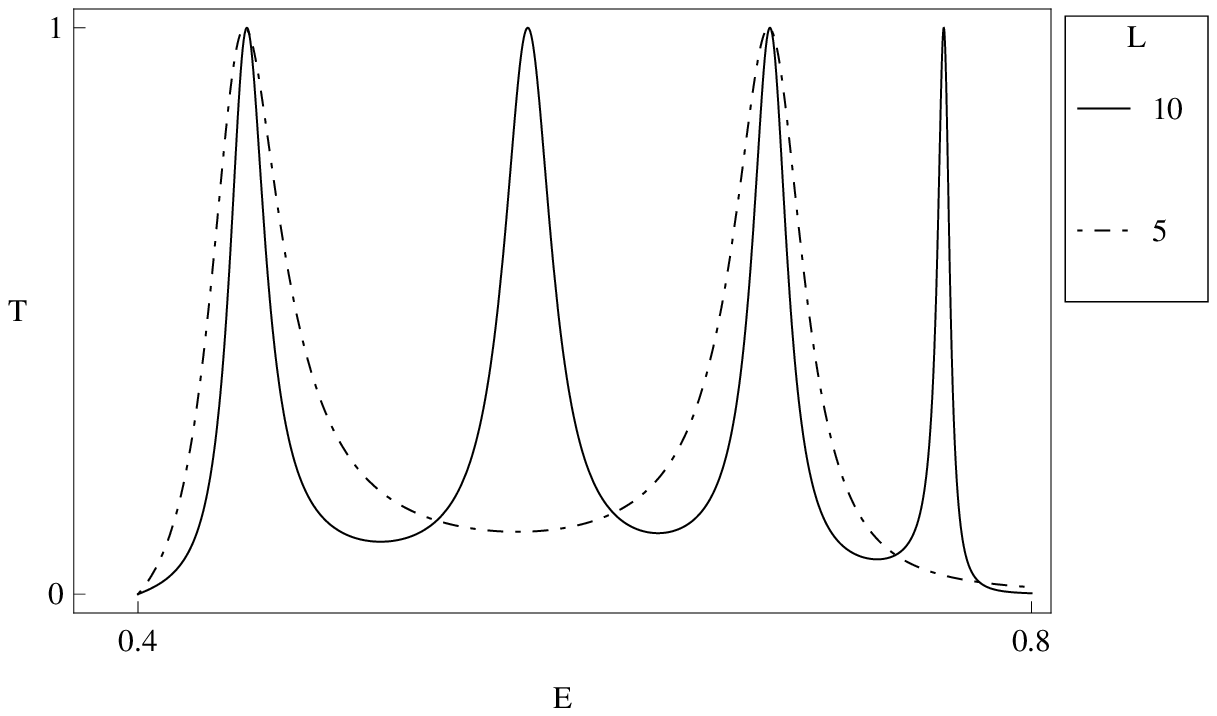}}
\caption{transmission coefficient versus energy $E$ with varying
$\alpha$ (left plot: $m_{0}=0.4, L=10, V_{0}=1.2$) and varying $L$
(right plot:$m_{0}=0.4, \alpha=5, V_{0}=1.2$) in case of PDM
($m_{1}=0.01$).}
\end{figure}

\begin{figure}
\centering \subfloat{
\includegraphics[width=2.8in]{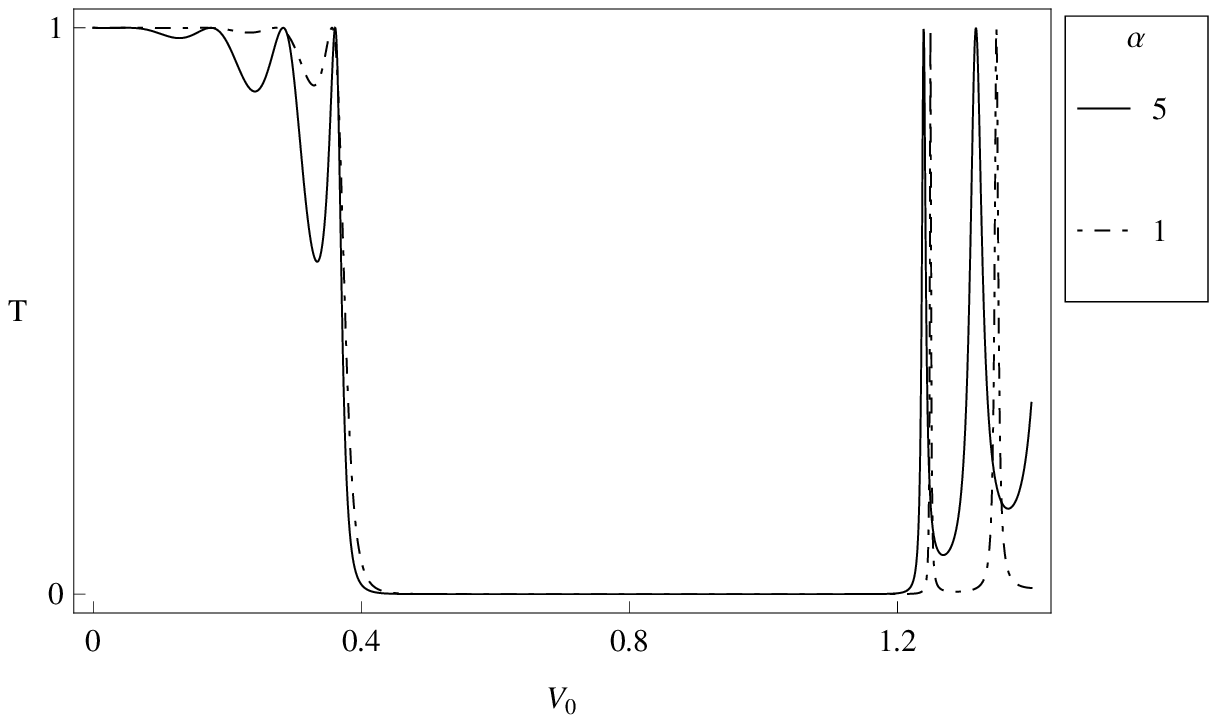}}
\hspace{0.1\linewidth} \subfloat{
\includegraphics[width=2.8in]{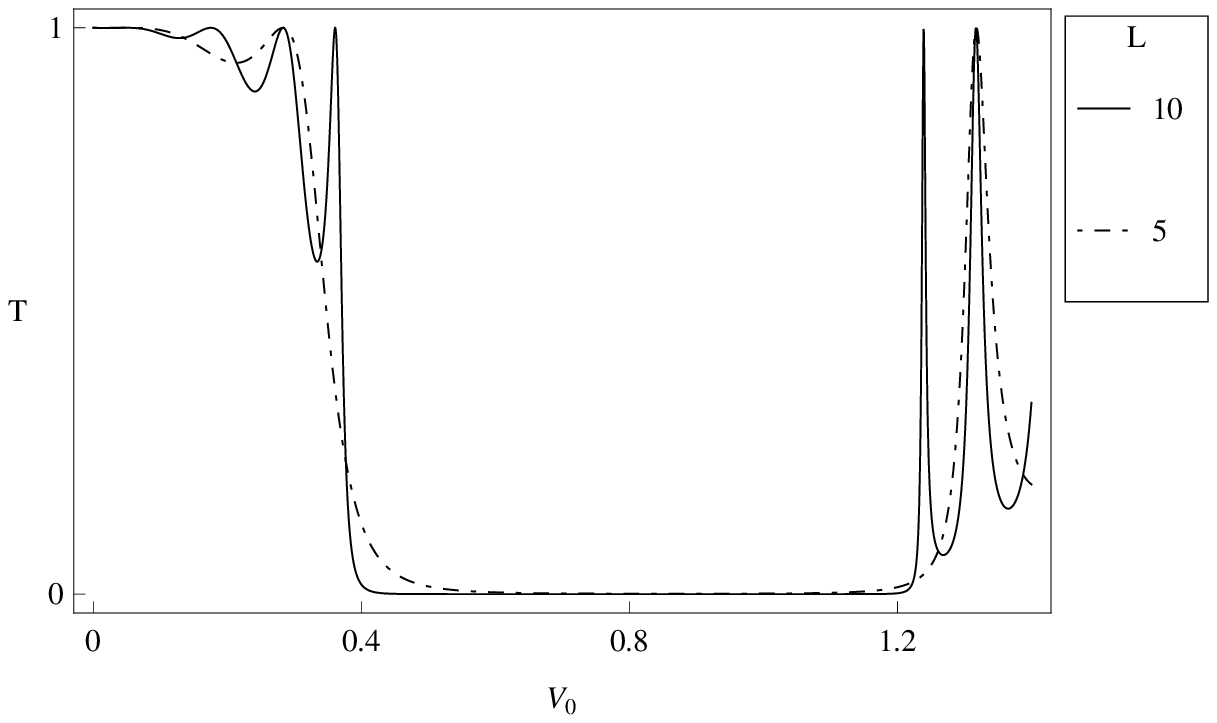}}
\caption{transmission coefficient versus potential parameter
$V_{0}$ with varying $\alpha$ (left plot: $m_{0}=0.4, L=10,
E=0.8$) and varying $L$ (right plot:$m_{0}=0.4, \alpha=5, E=0.8$)
in case of PDM ($m_{1}=0.01$).}
\end{figure}

\begin{figure}
\centering
\includegraphics[height=3in, width=4.5in, angle=0]{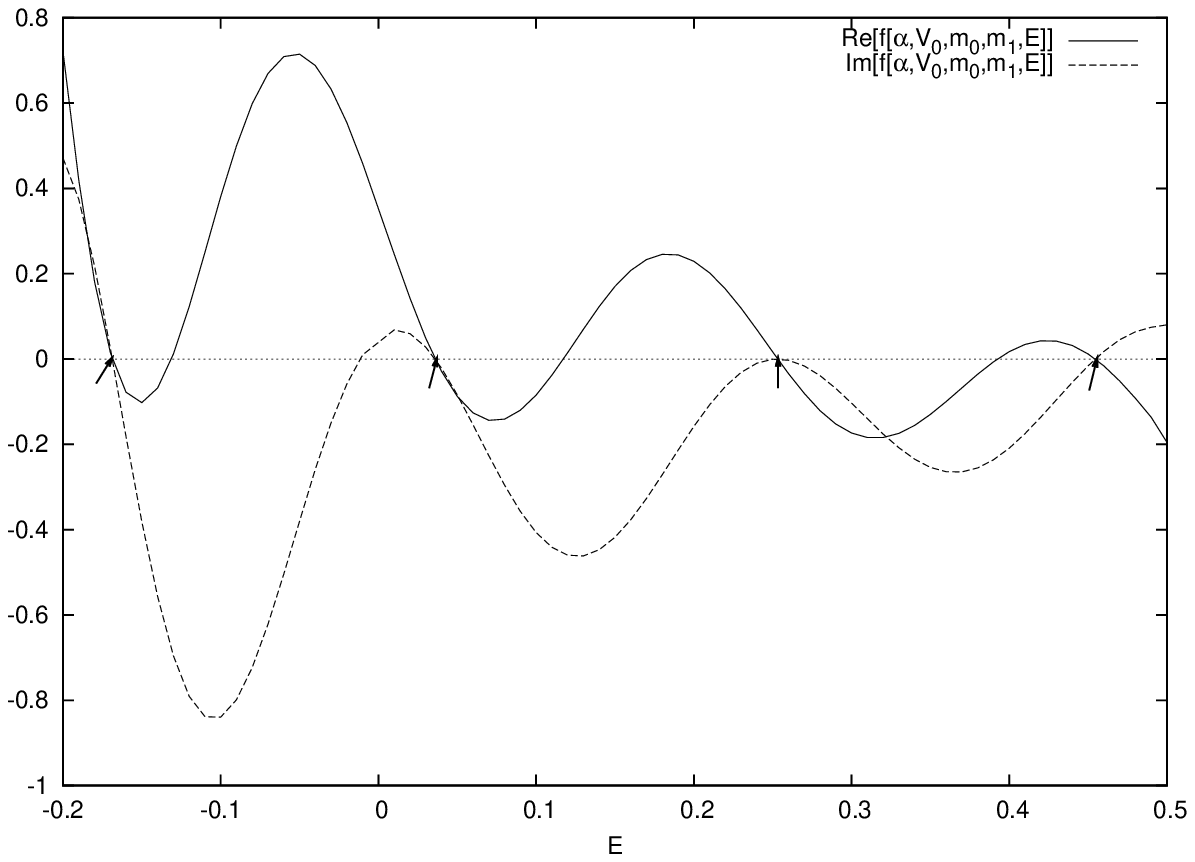}
\caption{plots of Re$[f[\alpha,V_{0},m_0,m_1,E]]$ and
Im$[f[\alpha,V_{0},m_0,m_1,E]]$  versus $E$ for $m_1=0.1$
($m_{0}=0.5, L=5, \alpha=10, V_{0}=1$).}
\end{figure}

\begin{figure}
\centering
\includegraphics[height=3in, width=4.5in, angle=0]{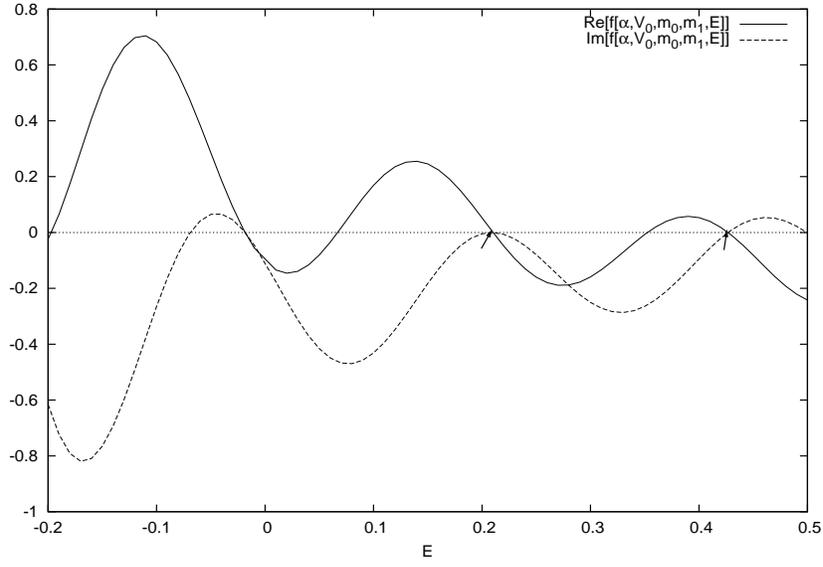}
\caption{plots of Re$[f[\alpha,V_{0},m_0,m_1,E]]$ and
Im$[f[\alpha,V_{0},m_0,m_1,E]]$  versus $E$ for the case of
constant mass ($m_{0}=0.5, L=5, \alpha=10, V_{0}=1$).}
\end{figure}

\end{document}